# From Graphene to Carbon Fibres: Mechanical Deformation and Development of a Universal Stress Sensor

**OTAKAR FRANK[1,†], GEORGIA TSOUKLERI[1,2], IBTSAM RIAZ[3], KONSTANTINOS PAPAGELIS[4], JOHN PARTHENIOS[1,2], ANDREA C. FERRARI[5], ANDRE K. GEIM[3], KOSTYA S. NOVOSELOV[3] AND COSTAS GALIOTIS[1,2,4*]**

[1]Institute of Chemical Engineering and High Temperature Chemical Processes, Foundation of Research and Technology-Hellas (FORTH/ICE-HT), Patras, Greece
[2]Interdepartmental Programme in Polymer Science and Technology, University of Patras, Patras, Greece
[3]Department of Physics and Astronomy, Manchester University, Manchester, UK
[4]Materials Science Department, University of Patras, Patras, Greece
[5]Engineering Department, Cambridge University, Cambridge, UK

**\*e-mail: c.galiotis@iceht.forth.gr**

[†] on leave from J. Heyrovsky Institute of Physical Chemistry, v.v.i., Academy of Sciences of the Czech Republic, Prague 8, Czech Republic

## Abstract

Carbon fibres (CF) represent a significant volume fraction of modern structural airframes. Embedded into polymer matrices, they provide significant strength and stiffness gains over unit weight as compared to other competing structural materials. Nevertheless, no conclusive structural model yet exists to account for their extraordinary properties. In particular, polyacrynonitrile (PAN) derived CF are known to be fully turbostratic: the graphene layers are slipped sideways relative to each other, which leads to an inter-graphene distance much greater than graphite. Here, we demonstrate that CF derive their mechanical properties from those of graphene itself. By monitoring the Raman G peak shift with strain for both CF and graphene, we develop a universal master plot relating the G peak strain sensitivity of all types of CF to graphene over a wide range of tensile moduli. A universal value of- average- shift rate with axial stress of $\sim -5\omega_0^{-1}(\mathrm{cm}^{-1}\mathrm{MPa}^{-1})$ is calculated for both graphene and all CF exhibiting annular ("onion-skin") morphology.

## Introduction

Graphene is a one-atom-thick planar sheet of sp$^2$-bonded carbon atoms that are densely packed in a two dimensional arrangement that forms a honeycomb crystal lattice. Graphene is the basic structural element of a number of carbon allotropes including graphite, carbon nanotubes and fullerenes. It can also be considered as an infinitely large aromatic molecule. Graphite, on the other hand, consists of many graphene sheets stacked together by out-of-plane (van der Waals) bonding forces. Other graphitic materials, such as carbon fibres or multiwall carbon nanotubes, are also governed, in terms of their physico-mechanical properties, by the extreme anisotropy between the strong intramolecular covalent forces and the weak intermolecular (van der Waals) forces. In carbon fibres (CF), the basic structural units are thought to be planar graphite crystals, with a lateral extension $L_a$ of nm dimensions and a stacking height $L_c$ of same order of magnitude[1]. However, this view has been challenged[2], based to the fact that the interplanar distance differs considerably from graphite. Single-wall carbon nanotubes (SWNT) can also be thought of as derived from the folding of graphene planes, whereas the morphology of multiwall carbon nanotubes (MWNT) has both graphene-tubular (longitudinal direction) and graphitic (through thickness direction) characteristics. Therefore, the detailed knowledge of the mechanical behaviour of the basic graphene structural unit should yield important information on the properties of graphitic structures, such as graphite itself and carbon fibres, but also of tubular structures at least in the longitudinal direction. Both planar and tube-like structures possess extraordinary tensile properties such as axial stiffness around 1 TPa [3] and tensile strengths varying from 2 GPa (high modulus CF)[4] or 9 GPa (CNT fibres)[5] to 50-60 GPa for individual SWNT[6] and MWNT[7], respectively.

The internal structure of carbon fibres has been the subject of numerous studies and as yet there is no model that describes perfectly well their complex texture and structure. For PAN-based carbon fibres, the prevalent notion is the presence of a basic structural unit (BSU) which governs their mechanical properties[1]. However, even for PAN-based fibres there is no agreement as to what the BSU consists of: this has been proposed to range from elongated ribbons[8] to even crumpled and folded sheets[9]. For graphitisation temperatures of less than 2500 °C [1] the packing in either the ribbons or sheets consists in turbostratic graphene layers, as the interlayer distance is markedly different than graphite. From the 002 reflection, Ref. [10] estimated 0.345-0.350nm for high-modulus fibres from different manufacturers and 0.355-0.360 nm for the high strength fibres. The situation is still more complicated for mesophase pitch fibres (MPP)[1], due to the variations in their processing routes that give rise to a number of textures ranging from radial to onion-like[1]. High brilliance synchrotron X-rays have been employed [2] to determine the nanostructure and texture of single carbon fibres of either PAN or MPP precursors: the basic structure is considered as composed of graphene layers of characteristic interplanar spacing for each fibre which is markedly greater than the characteristic value of graphite of 0.335 nm [9].

**Characterisation of graphene and carbon fibres with Raman Spectroscopy**

Since four decades ago a substantial amount of research has concentrated on the study of stress induced alterations in the Raman spectra of a whole range of materials both organic and inorganic[11-13]. In general, the vibrational spectra of crystalline materials are very sensitive to slight changes in their local molecular structure. These are usually changes in bond lengths, bond (valence) angles, internal rotation angles, etc. In graphitic materials, such as CF, the variation of phonon frequency per unit of strain can provide information on the efficiency of stress transfer to individual bonds[4]. This is because when a macroscopic stress is applied to a polycrystalline CF, the resulting deformation emanates not only from bond stretching or contraction, but also from a number of other mechanisms such as crystallite rotation and slippage, which do not change the phonon frequency. Indeed, the higher the crystalline order of a fibre (and hence the modulus[1]) the higher the degree of bond deformation and, hence, the higher the measured Raman shift per unit strain[4].

In all graphitic materials, the G peak corresponds to the doubly degenerate $E_{2g}$ phonon at the Brillouin zone centre. The D peak is due to the breathing modes of sp$^2$ rings and requires a defect for its activation[14]. It comes from TO phonons around the **K** point of the Brillouin zone, is active by double resonance (DR)[15] and is strongly dispersive with excitation energy due to a Kohn Anomaly at **K** [16]. DR can also happen as an intravalley process, i.e., connecting two points belonging to the same cone around **K** or **K**'. This gives rise to the so called D' peak, which can be seen around 1620 cm$^{-1}$ in defected graphite and CFs. High quality graphene shows the G, 2D, 2D' peaks, but negligible D, D' [17].

Previous Raman work on carbon fibres revealed that the G peak shifts are linearly related to macroscopically applied uniaxial stress for various families of quite diverse elastic moduli[4]. Indeed, it was reported that $\partial\omega/\partial\varepsilon$ is linearly related to fibre modulus, *E*, or, in other words, the $\partial\omega/\partial\sigma$ is a constant value[4,18], where $\omega$ is the wavenumber, $\varepsilon$ the strain and $\sigma$ axial stress in the fibre. Since the CF BSU[1] is different for each family member, it was concluded that the carbon fibres can be considered as equal-stress bodies[19]. The question that remains to be addressed is why $\partial\omega/\partial\sigma$ may vary when changes are brought about to the fibre precursor[18] and/or to the extent of fibre drawing within the carbonisation or stabilisation [4,18] regimes (regardless of the ultimate graphitisation "firing" temperature required for the development of graphite crystals). For example, for PAN-based fibres distinct values of $\partial\omega/\partial\sigma$ ranging

between –2.0 to –3.0 cm$^{-1}$ GPa$^{-1}$ are obtained for specific changes in the drawing procedure[4]. Also if one switches to a pitch-based precursor, instead of a PAN fibre, a straight line between $\partial\omega/\partial\varepsilon$ and $E$ of slope –1.6 cm$^{-1}$ GPa$^{-1}$ (i.e. the rate of shift per stress) is obtained[18]. However, in all the above works, the $\partial\omega/\partial\sigma$ has been estimated through direct $\partial\omega/\partial\varepsilon$ measurements divided by the nominal tensile modulus of the fibre. Since sampling by means of Raman scattering in graphitic materials is restricted to a depth of approximately 13-15 nm [20-23], there is uncertainty as to the correct use of $E$, which should correspond to the sampling volume (optical skin) of the fibre. As pointed out in Ref.[18], for fibres exhibiting skin-core differentiation this would lead to the suppression of the $\partial\omega/\partial\sigma$ values, since the effective "skin" modulus, which is in the denominator, may have a higher value than the nominal fibre modulus (re: for laser polarisation along the fibre axis). However, this cannot be the only source of discrepancy between the reported $\partial\omega/\partial\sigma$ values between PAN and pitch fibres since, direct measurements reported in the literature for a PAN fibre [24] yield a higher value than those derived in Ref.[18] for approximately the same modulus.

Here, we attempt to throw light into the phonon deformation of CF through the phonon shift of monolayer graphene itself. Mechanical experiments on monolayer graphene have been performed by bending poly(methyl methacrylate) (PMMA) cantilever beams in either four-point-bending (FPB)[25] or cantilever-beam configurations (CB)[26]. As demonstrated in Ref. [25], the imposition of a uniaxial stress field leads to the lifting of the degeneracy of $E_{2g}$ phonon and the splitting of the G peak into the G$^-$ and G$^+$ components. The eigenvectors of G$^-$ and G$^+$ peaks are parallel and perpendicular to the direction of the applied strain, respectively, and each has a different $\partial\omega/\partial\varepsilon$. Their relative Raman intensities are given by[25]:

$$I(G^-) \propto \sin^2(\theta_{in} + \theta_{out} + 3\varphi),\ I(G^+) \propto \cos^2(\theta_{in} + \theta_{out} + 3\varphi) \tag{1}$$

where $\varphi$ is the angle between the strain axis and the $x$ axis, chosen to be perpendicular to the C-C bond (Fig.1), $\theta_{in}$ and $\theta_{out}$ are the polarization of the incident and scattered light, for light incident perpendicular to the graphene plane, relative to the strain axis[25]. A question that now needs be addressed is whether the graphene orientation in carbon fibres also leads to observable phonon splitting and resulting differentiation of phonon stress/ strain shifts. This could affect the use of CF as potential stress/ strain sensors- in tandem to their reinforcing role- in commercial composites. Confirmation of the above paves also the way for determining the average orientation of the graphene units vis-à-vis the fibre axis (e.g. armchair, chiral or zigzag) information that, thus far, cannot be obtained by any other technique and can lead to the development of new CF fibres with tailor-made characteristics, that exploit not only the development of order but also the graphene orientation on the fibre surface.

The average orientation of the nano-BSU building block of CF with respect to the incident light is fixed during the carbonization and drawing regimes[1]. Therefore polarized Raman measurements could in principle determine the relative contribution of each G peak component. In contrast, high firing (graphitization) temperatures should not affect the $G^-/G^+$ intensity ratio since at such temperatures graphite microcrystals only grown at fixed orientation[1]. Indeed, it was long known that in CF the full width at half maximum of the G peak (FWHM(G)) increases with tensile strain[27], in contrast with the expectation that this should decrease following the decrease of crystallite misorientation and number of defects. We now interpret this as an indication of $E_{2g}$ phonon splitting during mechanical loading. We also note that strain itself has a negligible influence on FWHM(G), since this is a higher order effect, requiring much larger strains than found in experiments to be detectable[25]. However, it is worth noting here that, as shown in Fig.1, the graphene layers particularly in MPP fibres are not necessarily perpendicular to the incident beam and, therefore, the graphene orientation angle, $\varphi$, projected to the surface of the fibre diminishes as the graphene stack rotates in its axis.

Here we test two PAN-based and two pitch-derived fibres of various elastic moduli performing polarized Raman measurement with $\theta_{in}=0^0$ or $90^0$, and $\theta_{out}$ always set to $0^0$, i.e. incident light either parallel or perpendicular to the uniaxial strain axis (the fibre axis) and the scattered light collected placing and analyser aligned with the strain/fibre axis. The Raman shifts measured for these fibres are compared with those from a variety of PAN-based and pitch-based fibres reported earlier[18,27]. We are also revisit the analytical treatments developed for the stress sensitivity of the $E_{2g}$ phonon in graphite[28] and the corresponding strain sensitivity of same mode in graphene[25,28-30] in an attempt to unify both approaches and develop a universal formula for the linear dependence of both graphene and carbon fibres upon the imposition of a normal stress.

*Strain/ stress sensitivity of the G peak of graphene*

Fig. 2a plots a representative graphene Raman spectrum in the G peak spectral region. Fig. 2b shows the fitted G peak positions (Pos(G)) as a function of strain for a graphene monolayer loaded in a cantilever-beam (CB) configuration. Care was taken to measure graphene flakes that exhibited zero residual strain in order to avoid the occurrence of non-linear $\Delta\omega$ vs. strain effects[25]. The initial stress condition of the flake depends on fabrication (mechanical cleavage) and/or surrounding film shrinkage due to curing. The straight lines are least-squares-fitted fits out of four different sets of measurements that yield on average $\partial \omega_{G^+}/\partial \varepsilon = -9.7 \times 10^2$ cm$^{-1}$ and $\partial \omega_{G^-}/\partial \varepsilon = -31.5 \times 10^2$ cm$^{-1}$ for the two components of the G peak, in good agreement with previous measurements employing a different loading device[25]. To derive the phonon shifts with respect to axial strain for unsupported graphene and therefore eliminate any effect from the Poisson's ratio of substrate, we can first estimate the Gruneisen and the shear deformation potential parameters for the supported case as shown in [25] and then revert back to free graphene. Such procedure yields values of $\partial \omega_{G^+}/\partial \varepsilon = -17.5 \times 10^2$ cm$^{-1}$ and $\partial \omega_{G^-}/\partial \varepsilon = -36.0 \times 10^2$ cm$^{-1}$, which compare well with the values of $-18.6 \times 10^2$ cm$^{-1}$ and $-36.4 \times 10^2$ cm$^{-1}$, obtained earlier[25].

The secular equation for $E_{2g}$ mode of graphene under strain is (see *Supporting Information*) [25,28-30]:

$$\begin{vmatrix} A\varepsilon_{xx} + B\varepsilon_{yy} - \lambda & (A-B)\varepsilon_{xy} \\ (A-B)\varepsilon_{xy} & B\varepsilon_{xx} + A\varepsilon_{yy} - \lambda \end{vmatrix} = 0 \qquad (3)$$

where $\lambda = (\omega^2 - \omega_0^2) \approx 2\omega_0 \Delta\omega$ is the difference between the squared strain dependence frequency, $\omega$, and the squared frequency in the absence of strain, $\omega_0$.

Solving analytically equation (3) and ignoring terms higher that $\varepsilon^2$, the G$^-$ and G$^+$ shifts in uniaxially strained graphene in *x* direction ($\varepsilon_{xx} = \varepsilon, \varepsilon_{yy} = -\nu\varepsilon, \varepsilon_{xy} = 0$) are given by:

$$\frac{\partial \omega_{G^+}}{\partial \varepsilon} = \frac{B - \nu A}{2\omega_0} \qquad (4)$$

$$\frac{\partial \omega_{G^-}}{\partial \varepsilon} = \frac{A - \nu B}{2\omega_0} \qquad (5)$$

or

$$A = \frac{2\omega_0 \left( \dfrac{\partial \omega_{G^-}}{\partial \varepsilon} + \nu \dfrac{\partial \omega_{G^+}}{\partial \varepsilon} \right)}{1 - \nu^2} \qquad (6)$$

$$B = 2\omega_0 \frac{\partial \omega_{G^+}}{\partial \varepsilon} + \nu A \qquad (7)$$

By inserting the estimated values of $\partial \omega_{G^+}/\partial \varepsilon$ and $\partial \omega_{G^-}/\partial \varepsilon$ for free graphene in equations (6) and (7), we obtain $A = -1.23 \times 10^7 \, \text{cm}^{-2}$ and $B = -7.16 \times 10^6 \, \text{cm}^{-2}$ which conform well with those inferred in Ref [25,30].

We now consider the stress sensitivity. For uniaxial stress in graphene, the resulting strains are given by $\varepsilon_{xx} = \varepsilon = S_{11}\sigma$ and $\varepsilon_{yy} = -\nu\varepsilon = S_{12}\sigma$ or equivalently

$$\varepsilon = S_{11}\sigma \tag{8}$$

$$\nu = -\frac{S_{12}\sigma}{\varepsilon} \tag{9}$$

where $S_{11}$ and $S_{12}$ are the compliances. Following the analysis mentioned earlier we obtain:

$$\frac{\partial \omega_{G^+}}{\partial \sigma} = \frac{(AS_{12} + BS_{11})}{2\omega_0} \tag{10}$$

$$\frac{\partial \omega_{G^-}}{\partial \sigma} = \frac{(AS_{11} + BS_{12})}{2\omega_0} \tag{11}$$

In order to derive the stress sensitivity of the G peak we need to know the compliance constants for graphene. From $S_{11}^{graphene} = 1/E_{11}^{graphene} = 1.00 \, \text{TPa}^{-1}$ and $S_{12}^{graphene} = 1/E_{12}^{graphite} = -0.16 \, \text{TPa}^{-1}$ [31] we obtain $\partial \omega_{G^+}/\partial \sigma = -1.6 \, \text{cm}^{-1}\text{GPa}^{-1}$ and $\partial \omega_{G^-}/\partial \sigma = -3.5 \, \text{cm}^{-1}\text{GPa}^{-1}$ [32]. The corresponding values from experiments on an embedded flake using the CB apparatus[26] are $\sim -1.7$ and $\sim -3.6 \, \text{cm}^{-1} \, \text{GPa}^{-1}$, respectively, whereas those derived earlier on a bare flake using the FPB test[25] were $-1.9$ and $-3.6 \, \text{cm}^{-1} \, \text{GPa}^{-1}$. The agreement between the above analysis and both experimental tests is indeed excellent and does not depend on the type of polymer beams (CB or FPB) employed to load the monolayer.

*Strain/ Stress sensitivity of the G-peak of carbon fibres*

A polycrystalline (turbostatic) CF under uniaxial tension can be seen as stacks of graphene. In this case both G$^-$ and G$^+$ peaks should be seen. As shown in[25], only for specific combinations of incident polarisation/ analysis directions one of the two components can be suppressed. In almost all literature on carbon fibres (see for example Refs. [4,18,27]) the incident polarisation is along the fibre ($\theta_{in}=0^0$) but no analyser is used, and no G peak splitting is seen, but an increase of FWHM(G). We assign the observed large increase of FWHM(G) as a signature of to phonon splitting and associated differences in the relative shift rate with strain. To test this hypothesis we subject our fibres to tension and examined in detail the peak evolution with strain, for $\theta_{in}=0^0$ or $90^0$ and $\theta_{out}=0^0$.

For the high-modulus (HM) PAN based fibre the G peak is plotted in Fig.3a at 0 and 1% strain. For strains higher than 0.5-0.6% the fitting of the G-peak with two Lorentzian bands (black line) is better than a single Lorentzian fit (red line, Fig.3a) (see Materials and Methods). The two components correspond to $G^-$ and $G^+$ as shown in Fig.2a. For a tensile strain of 1% and $\theta_{in}=\theta_{out}=0^0$, we get $I(G^-)/I(G^+) \sim 0.6$ from a number of fibres tested (Fig.3a). From Eq. 1, and $\theta_{in}=\theta_{out}=0^0$, we expect $I(G^-)/I(G^+) \propto \tan^2(3\varphi)$, implying an average $\varphi \sim 13^0$ (Fig.1). However, this result should be treated with caution even for "onion" skin type of fibres due to the presence of disorder that can affect the measured intensities. In contrast, for the P55 fibre there is no sign of splitting at strains just prior to fracture (~0.8%) indicating as argued below, the absence of the $G^-$ component in MPP fibres.

In Fig.4, the position of the G peak [Pos(G)] is plotted with strain for both the HM and P55 fibres. For the HM fibre (annular morphology), FWHM(G) increases significantly for strains higher than 0.4% whereas the P55 fibre shows no increase up to fracture. Also shown in Fig.4, is the position of the G peak for both $\theta_{in}=0^0$ and $\theta_{in}=90^0$ and ($\theta_{out}=0$). The $\partial\omega_G/\partial\varepsilon$ for all fibres tested is $-11.2\times10^2\,\text{cm}^{-1}$ for $\theta_{in}=0^0$ up to 0.4%, similar to what reported earlier for similar fibres[18,27] and $-9.5\times10^2\,\text{cm}^{-1}$ for $\theta_{in}=90^0$. However, at higher strains by fitting two Lorentzians, we get $\partial\omega_{G-}/\partial\varepsilon$ and $\partial\omega_{G+}/\partial\varepsilon$ of $-18.8\times10^2\,\text{cm}^{-1}$ and $-6.7\times10^2\,\text{cm}^{-1}$ (Fig.4a). In contrast, the P55 fibre (Fig.4b), with radial morphology[18], shows no appreciable FWHM(G) increase, hence no sign of peak splitting. It is interesting to note that the average $\partial\omega_G/\partial\varepsilon$ for the P55 is very close to that of the $G^+$ peak in the HM fibre of similar modulus. Translating this finding to the graphene plane (see Fig.5 in [25]), prevalence of $G^+$ would imply that $\varphi$ is close to zero, i.e. the C-C bond is perpendicular to the fibre axis (armchair configuration).

The expected shift per units of normal stress in fibres can be derived as follows. If both components contribute equally to the measured Raman shift, Eq(6) becomes:

$$\Delta\omega_G = \left(\frac{\Delta\omega_{G^+} + \Delta\omega_{G^-}}{2}\right) = \left[\frac{(A+B)(S_{11}+S_{12})}{4\omega_0}\right]\sigma \approx \left[\frac{S_{11}(A+B)}{4\omega_0}\right]\sigma \qquad (12)$$

By returning to the treatment presented in the previous section for graphene, we can estimate the theoretical value of the expression in brackets as $\sim -5\omega_0^{-1}(\text{cm}^{-1}\text{MPa}^{-1})$. This value has a universal validity at least for fibres exhibiting annular morphology (most PAN-based fibres) and confirms the applicability of fibre as stress sensors in a number of applications. As shown below, it also defines the mean value of phonon shift over a wide modulus range.

**The universal master plot**

In Fig. 5, all $\partial\omega_G/\partial\varepsilon$ obtained for graphene, the PAN-based, HM and IM fibres, as well as, the MPP-based, P55 and P25 fibres, are plotted against their nominal- bulk- tensile Young's modulus. All these data were obtained for $\theta_{in}=0^0$ and $\theta_{out}=0^0$. Furthermore we include on the same graph data from PAN-based and MPP fibres obtained over the last 20 years from Refs. [27], [18], respectively (see Materials and Methods for details of these fibres). As shown, all CF data are contained within the two boundary lines defined by the $G^+$ and $G^-$ slopes of graphene of $\sim -1.7$ and $\sim -3.6$ cm$^{-1}$ GPa$^{-1}$, respectively. The theoretical boundaries are also shown on the same graph and are close to the above values. The above confirms the CF affinity – in terms of their mechanical response – to their fundamental building unit, which is graphene itself. It is thus more appropriate to term carbon fibres as "graphene stacks" rather than "graphite fibres" which was the prevalent notion for decades. Certain effects observed in Fig. 5 are also worth commenting upon.

The data points from fibres exhibiting annular structure (classical processing route[1]) lie below the bisector (average) line exhibiting a $\partial\omega_G/\partial\sigma = -2.7$ cm$^{-1}$GPa$^{-1}$, which has also been confirmed by direct stress measurements[24]. On the contrary, PAN-based fibres produced by the imposition of a higher drawing ratio at the carbonization stage and at relative lower ultimate firing (graphitisation) temperatures (UFT) exhibit a marked decrease of $\partial\omega_G/\partial\sigma$. The conclusive proof for this phenomenon is given for the so-called Group C fibres[27] of approximately the same modulus but of varying draw ratio; as can be seen the higher the extend of pre-graphitisation drawing and thus the lower the UFT, the smaller the value of obtained $\partial\omega_G/\partial\sigma$. It appears therefore that as the morphology of the fibre is altered by drawing from the annular ("onion") to a highly folded structure[27], the data points shift towards the $G^+$ boundary (see Fig. 5). Finally, the data for MPP fibres presented here and in Ref.[18], with radial morphology, tend to conform to the $G^+$ line of slope $\partial\omega_G/\partial\sigma = -1.6$ cm$^{-1}$GPa$^{-1}$. These effects can be explained by the results of the previous sections with reference to the angle, $\varphi$, between the strain axis and the C-C bond (Fig.1). As the graphene stacks in CF rotate and the $\varphi$ decreases towards $0^0$ (radial morphology, Fig.1) the $G^+$ peak prevails, for eq. (1) with $\theta_{in}=\theta_{out}=0^0$. Finally, the $G^-$ and $G^+$ peaks strain shifts of $-18.8\times10^{-2}\,\text{cm}^{-1}$ and $-6.7\times10^{-2}\,\text{cm}^{-1}$ of the HM fibre (Fig.4a), respectively, can be projected onto the average line of Fig.5 in order to get an estimate of the true tensile modulus of the optical (sampling) area of the fibre. The results indicate a skin modulus of some 100 GPa higher than the bulk modulus of the fibre which in broad agreement with the results in Ref.[18] by means of X-ray measurements for fibres of similar morphology.

**Conclusions**

The acquired knowledge of the G peak shift and splitting with uniaxial strain in graphene has been employed to interpret the mechanical response of carbon fibres of various types. It has been shown that, for polarised measurements, accurate determination of the G peak shift and splitting with stress reveal very important morphological issues that, thus far, have not been shown by any other technique. These are the identification of the average orientation of the graphene units (or stacks) vis-a-vis the fibre (strain) axis and, in certain cases, of the optical skin modulus. Furthermore, by comparing the results derived for graphene and carbon fibres, a universal plot has been constructed that relates the G peak shift to stress or strain for all graphitic materials. Both theory and experiment yield a universal value of- average-G peak shift with stress of $\sim -5\omega_0^{-1}(\text{cm}^{-1}\text{MPa}^{-1})$ regardless of modulus. In short, although the notion of "stress" is hard to define for a monolayer, its phonon stress derivative converts it into the most powerful non-evasive stress sensor for a plethora of applications.

*Materials and Methods*

*Carbon Fibres (this work)*

The HM fibre is 7 μm in diameter of bulk modulus of 370 GPa band is produced by a first generation manufacturing technology[27]. Acrylic filaments were wet-spun and drawn in hot water and saturated steam to a total draw ratio of 14 times, to yield a final diameter of approximately 12 μm. The filaments were stabilized in hot air until the density had risen to $1.38-1.40\,gcm^{-3}$, using a rising-ramped temperature regime (225-245 C). Primary carbonization was carried out using a maximum temperature of 950 C, whilst secondary carbonization and graphitization utilized ultimate firing temperatures (UFT) of 2600 C. No additional drawing processes were carried out. During the stabilization, the filaments were held at constant length, whilst during the carbonization and graphitization, a 5% shrinkage was allowed. The IM fibre consists of a 5 μm diameter fibre bulk modulus of 270 GPa. In this case, acrylic filaments were wet-spun and drawn as described above. The filaments were then subjected to a multi-stage pre-stabilization drawing at temperatures up to 270 C, followed by stabilization as described above. Carbonization and graphitization were carried out with the filaments held at constant length. The UFT value for the IM fibre is 1750 C.

Two commercial MPP fibres of 167 GPa (P25) and 371 GPa (P55) in tensile moduli, produced and supplied by Cytec Industries (US), were also tested. These carbon fibers are produced using mesophase pitches (MPP)[1]. Pitch, itself, is produced from petroleum or coal tar which is made up of fused aromatic rings. The production of pitch-based carbon fibers involves melt spinning of pitch precursor fibers, stabilization (oxidation), carbonization, and graphitization[1].

*Carbon Fibres (of Refs.[18,27])*

Group A and Group B fibres shown in the Master Plot of Fig.5 were also produced with the manufacturing processes mentioned above for HM and IM fibres. Group C consists of three approximately 6 μm diameter fibres of similar Young's modulus but of distinctly different morphologies[27]. Acrylic filaments were also treated as in Group A fibres but different drawing procedures were pursued in the stabilization (up to 270 C) and carbonization regimes (up to 950 C) so as the higher the drawn fibre the lower UFT that was required for the attainment of the 370 GPa modulus[27]. As shown in Fig. 5, by altering just the drawing procedure and reducing the UFT the $\partial \omega_G / \partial \sigma$ shifts gradually to lower values ($G^+$ dominance). The data from a whole range of MPP fibres reported in [18] were also shown in Fig. 5

*Graphene monolayers*

Graphene monolayers were prepared by mechanical cleavage from natural graphite (Nacional de Grafite) and transferred onto the PMMA cantilever beam covered by a ~200 nm thick layer of SU8 photoresist (SU8 2000.5, MicroChem). After placing the graphene samples, a thin layer of S1805 photoresist (Shipley) was spin-coated on the top. The beam has a total thickness of $t = 2.9$ mm and width $b = 12.0$ mm. The graphene flake was located at a distance, $x$, from the fixed end of 12.97 mm. The flake under study has dimensions of approximately 6x56 μm with the shorter side parallel to the strain axis. The details of the cantilever beam technique are given in *Supporting Information*.

*Raman Measurements*

MicroRaman (InVia Reflex, Renishaw, UK) spectra of graphene were recorded with 785 nm (1.58eV) excitation, while the laser power was kept below 0.85 mW to avoid heating. The high excitation wavelength was required to suppress the fluorescence of the polymer coating. A 100x objective with numerical aperture of 0.9 is used, and the spot size is estimated to be ~1x2 µm. The polarization of the incident light was kept parallel to the applied strain axis. Because the graphene peaks overlap with strong peaks originated from the substrate, the spectra were first baseline (linear) subtracted, then normalized to its most intense peak of the substrate at 1450 cm$^{-1}$, and subsequently the spectrum of bare substrate was subtracted. All bands in the Raman spectra of graphene were fitted with Lorentzians. The FWHM of the G peak for the unstressed graphene was found to be approximately 6-8 cm$^{-1}$.

MicroRaman spectra of carbon fibers were measured at 514.5 nm (2.41eV) with a laser power of below 1.1 mW. A 80x objective with numerical aperture of 0.75 is used, and the spot size is estimated to be ~1 µm. The data are collected in back-scattering and with a triple monochromator and a Peltier cooled CCD detector system. The polarization of the incident light was either parallel ($\theta_{in}=0^0$) or perpendicular ($\theta_{in}=90^0$) to the applied strain axis. The polarization of the scattered light was selected to $\theta_{out}=0^0$.

*Testing carbon fibres in air*

Individual carbon fibres in air were bonded to the jaws of a small straining with their axes aligned parallel to the stretching direction to $\pm 5^0$. The gauge lengths of the fibres fixed to 25.00 mm and the extension of the fibres could be measured to $\pm$ µm. The spectra were taken close to the middle of the fibre and for each step five measurements were made. In all cases, the intensity values in text refer to the integrated area of the respective Raman bands.

The HM carbon fibre $\partial\omega_G/\partial\varepsilon$ shift rate was determined from 8 independent experiments at incident laser polarization of $\theta_{in} = 0°$ and 5 experiments with $\theta_{in} = 90°$, using at least 3 measurements at every strain level. The strain was increased in steps 0.05 – 0.2 % up to failure at 0.8-0.11%. The spectra were using D, G and D' bands with Lorentzian line shapes. For $\theta_{in} = 0°$, the least-squares fits of Pos(G) vs. ε were performed on every experiment and then averaged giving $\partial\omega_G/\partial\varepsilon$ of -11.2 ± 0.7 x 10$^2$ cm$^{-1}$ as well as on all data giving $\partial\omega_G/\partial\varepsilon$ of -11.2 ± 0.3 x 10$^2$ cm$^{-1}$. For $\theta_{in} = 90°$, the least-squares fit of Pos(G) vs. ε was performed on all acquired data giving $\partial\omega_G/\partial\varepsilon$ of -9.5 ± 0.3 x 10$^2$ cm$^{-1}$. For the G peak deconvolution, 6 sets of experiments with longer accumulation times were attempted. The FWHM of the $G^+$ and $G^-$ sub-bands was set to be equal during the deconvolution and values between 28 and 29 cm$^{-1}$ were obtained consistently, which correspond to the FWHM of the fibres at zero strain. Due to statistically small differences of the $R^2$ value between fits using one or two Lorentzian lineshapes for the G peak, the quality of the fits was assessed also be the evolution of the F-values. For a particular experiment, no significant difference in F-values was observed until a strain level of approx. 0.5-0.6%. From this point onwards, the F-value for the single G peak fit was getting progressively lower in comparison to the $G^+$, $G^-$ fit.

The IM carbon fibre $\partial \omega_G / \partial \varepsilon$ shift rate was obtained from two independent experiments, using three measurements at every strain level. The strain was increased in steps < 0.1% up to failure at approx 0.9%. The spectra were fitted using D, D'', G and D' bands with Lorentzian line shapes. The least-squares line fit of Pos(G) vs. ε was performed on all acquired data to obtain the $\partial \omega_G / \partial \varepsilon$ rate of -6.2 ± 0.8 x 10$^2$ cm$^{-1}$.

The P25 carbon fibre $\partial \omega_G / \partial \varepsilon$ shift rate was obtained from two independent experiments, using three measurements at every strain level. The strain was increased in steps < 0.1% up to failure at approx 0.75%. The spectra were fitted using D, D'', G and D' bands with Lorentzian line shapes. The least-squares line fit of Pos(G) vs. ε was performed on all acquired data to obtain the $\partial \omega_G / \partial \varepsilon$ rate of -2.15 ± 0.86 x 10$^2$ cm$^{-1}$ (the value after ± is the 95% confidence interval).

The P55 carbon fibre $\partial \omega_G / \partial \varepsilon$ shift rate was determined from 5 independent experiments, using at least 3 measurements at every strain level. The strain was increased in steps < 0.1% up to failure at 0.55-0.8%. The spectra were fitted using using D, G and D' bands with Lorentzian line shapes. The least-squares fits of Pos(G) vs. ε were performed on every experiment and then averaged giving $\partial \omega_G / \partial \varepsilon$ of -5.26 ± 1.3 x 10$^2$ cm$^{-1}$ as well as on all data giving $\partial \omega_G / \partial \varepsilon$ of -5.28 ± 1.6 x 10$^2$ cm$^{-1}$.

**Author contributions**

Project planning, C.G.; sample preparation G.T., I.R., K.S.N., A.K.G.; spectroscopic measurements and analysis, G.T., O.F., J.P, K.P., N.M., A.C.F.; data interpretation O.F., K.P., C.G, A.C.F.; manuscript drafting, C.G., K.P., O.F., A.C.F.**Acknowledgements**

FORTH/ICE-HT acknowledges financial support from the Marie-Curie Transfer of Knowledge program CNTCOMP [Contract No.: MTKD-CT-2005-029876]. Also, GT gratefully acknowledges FORTH/ICE-HT for a scholarship and ACF, KN, AKG thank the Royal Society and the European Research Council for financial support. ACF acknowledges funding from EPSRC grant EP/G042357/1 Cytec Industries (US) are thanked for supplying the P25 and P55 fibres.

**REFERENCES**

1.  Morgan, P. *Carbon fibers and their composites*. (Taylor & Francis, Boca Raton, 2005).

2.  Loidl, D., Peterlik, H., Muller, M., Riekel, C., & Paris, O. Elastic moduli of nanocrystallites in carbon fibers measured by in-situ X-ray microbeam diffraction. *Carbon* 41 (3), 563-570 (2003).

3.  Treacy, M.M.J., Ebbesen, T.W., & Gibson, J.M. Exceptionally high Young's modulus observed for individual carbon nanotubes. *Nature* 381 (6584), 678-680 (1996).

4.  Melanitis, N., Tetlow, P.L., Galiotis, C., & Smith, S.B. Compressional Behavior of Carbon-Fibers .2. Modulus Softening. *J. Mater. Sci.* 29 (3), 786-799 (1994).

5.  Koziol, K. *et al.* High-performance carbon nanotube fiber. *Science* 318 (5858), 1892-1895 (2007).

6.  Yu, M.F., Files, B.S., Arepalli, S., & Ruoff, R.S. Tensile loading of ropes of single wall carbon nanotubes and their mechanical properties. *Phys. Rev. Lett.* 84 (24), 5552-5555 (2000).

7.  Yu, M.F. *et al.* Strength and breaking mechanism of multiwalled carbon nanotubes under tensile load. *Science* 287 (5453), 637-640 (2000).

8.  Ruland, W. Carbon-Fibers. *Adv. Mater.* 2 (11), 528-536 (1990).

9.  Oberlin, A. Carbonization and graphitization. *Carbon* 22 (6), 521-541 (1984).

10. Guigon, M. & Oberlin, A. Heat-treatment of high tensile strength PAN-based carbon fibres: Microtexture, structure and mechanical properties. *Compos. Sci. Technol.* 27 (1), 1-23 (1986).

11. Anastassakis, E., Pinczuk, A., Burstein, E., Pollak, F.H., & Cardona, M. Effect of Static Uniaxial Stress on Raman Spectrum of Silicon. *Solid State Commun.* 8 (2), 133-& (1970).

12. Cerdeira, F., Buchenauer, C.J., Cardona, M., & Pollak, F.H. Stress-Induced Shifts of First-Order Raman Frequencies of Diamond and Zinc-Blende-Type Semiconductors. *Phys. Rev. B* 5 (2), 580-& (1972).

13. Guild, F.J., Vlattas, C., & Galiotis, C. Modeling of Stress Transfer in Fiber Composites. *Compos. Sci. Technol.* 50 (3), 319-332 (1994).

14. Ferrari, A.C. & Robertson, J. Interpretation of Raman spectra of disordered and amorphous carbon. *Phys. Rev. B* 61 (20), 14095-14107 (2000).

15. Thomsen, C. & Reich, S. Double resonant Raman scattering in graphite. *Phys. Rev. Lett.* 85 (24), 5214-5217 (2000).

16. Piscanec, S., Lazzeri, M., Mauri, F., Ferrari, A.C., & Robertson, J. Kohn anomalies and electron-phonon interactions in graphite. *Phys. Rev. Lett.* 93 (18) (2004).

17. Ferrari, A.C. *et al.* Raman spectrum of graphene and graphene layers. *Phys. Rev. Lett.* 97 (18), 187401 (2006).

18. Huang, Y. & Young, R.J. Effect of fibre microstructure upon the modulus of PAN- and pitch-based carbon fibres. *Carbon* 33 (2), 97-107 (1995).

19. Northolt, M.G., Veldhuizen, L.H., & Jansen, H. Tensile Deformation of Carbon-Fibers and the Relationship with the Modulus for Shear between the Basal Planes. *Carbon* 29 (8), 1267-1279 (1991).

20. Casiraghi, C. *et al.* Rayleigh imaging of graphene and graphene layers. *Nano Lett.* 7 (9), 2711-2717 (2007).

21. Kravets, V.G. *et al.* Spectroscopic ellipsometry of graphene and an exciton-shifted van Hove peak in absorption. *Phys. Rev. B* 81 (15), 155413 (2010).

22. Wada, N. & Solin, S.A. Raman efficiency measurements of graphite. *Physica B+C* 105 (1-3), 353-356 (1981).

23. Djurisic, A.B. & Li, E.H. Optical properties of graphite. *J. Appl. Phys.* 85 (10), 7404-7410 (1999).

24. Chohan, V. & Galiotis, C. Effects of interface, volume fraction and geometry on stress redistribution in polymer composites under tension. *Compos. Sci. Technol.* 57 (8), 1089-1101 (1997).

25. Mohiuddin, T.M.G. *et al.* Uniaxial strain in graphene by Raman spectroscopy: G peak splitting, Grueneisen parameters, and sample orientation. *Phys. Rev. B* 79 (20), 205433-205438 (2009).

26. Tsoukleri, G. *et al.* Subjecting a Graphene Monolayer to Tension and Compression. *Small* 5 (21), 2397-2402 (2009).

27. Melanitis, N., Tetlow, P.L., & Galiotis, C. Characterization of PAN-based carbon fibres with laser Raman spectroscopy .1. Effect of processing variables on Raman band profiles. *J. Mater. Sci.* 31 (4), 851-860 (1996).

28. Sakata, H., Dresselhaus, G., Dresselhaus, M.S., & Endo, M. Effect of Uniaxial-Stress on the Raman-Spectra of Graphite Fibers. *J. Appl. Phys.* 63 (8), 2769-2772 (1988).

29. Huang, M.Y. *et al.* Phonon softening and crystallographic orientation of strained graphene studied by Raman spectroscopy. *Proc. Natl. Acad. Sci. U.S.A.* 106 (18), 7304-7308 (2009).

30. Thomsen, C., Reich, S., & Ordejon, P. Ab initio determination of the phonon deformation potentials of graphene. *Phys. Rev. B* 65 (7), 073403 (2002).

31. Lee, C., Wei, X.D., Kysar, J.W., & Hone, J. Measurement of the elastic properties and intrinsic strength of monolayer graphene. *Science* 321 (5887), 385-388 (2008).

32. Blakslee, O.L., Proctor, D.G., Seldin, E.J., Spence, G.B., & Weng, T. Elastic Constants of Compression-Annealed Pyrolytic Graphite. *J. Appl. Phys.* 41 (8), 3373-3382 (1970).

**FIGURE CAPTIONS**
1. Schematic representation of scattering effects in the plane of graphene for onion-skin (left) and radial (right) carbon fibres. $\vec{\varepsilon}$ depicts the strain axis whereas $e_i$, and $e_s$ *are the* polarization directions of the incident and scattered light, respectively. $\theta_{in}$ and $\theta_{out}$ are the angles between the strain axis and the plane of the electric-field vector of incident and scattered light, respectively. The *x* axis has been taken perpendicular to the C-C bond. $\varphi$ is the angle between the strain axis and the *x* axis (i.e. orientation of the graphene lattice with respect to strain). $L_c$ is the carbon fibre crystallite thickness (i.e. number of graphene layers). $L_{a//}$ and $L_{a\perp}$ represent crystallite width in directions parallel and perpendicular to the fiber axis, respectively. $e_i$ is plotted in both directions, which were used in the experiments ($\theta_{in}$ = 0° and 90°). In the bottom right panel $e_{ip}$ designates the $e_i$ vector projected onto the graphene plane, which is rotated around a vertical axis at an arbitrary angle. $e_s$ is plotted only at $\theta_{out}$ = 0°.

2. a) G peak Raman spectra of graphene at 0% (grey) and 1% (black) strain at 785 nm excitation. The original measurements are plotted as points. The solid curves are the best Lorentzian fits to the experimental spectra. b) $G^+$ (empty rectangles) and $G^-$ (full rectangles) peak positions for graphene embedded in polymer beam in tension. The data points are averages from five different locations on the flake. Solid lines represent values calculated for suspended graphene in air (see text). The inset shows an optical microphotograph of the measured flake. The scale bar on the inset is 10 μm and the arrows indicate the strain orientation.

3. Raman G band region of a) HM fibre at 0 (G+D', grey) and 1% ($G^+$+$G^-$+D', black) strain and b) of P55 fibre at 0 (G+D', grey) and 0.8% (G+D', black) strain. The original measurements are plotted as points. The lines of respective colors represent best Lorentzian fits to the experimental spectra, where the dashed lines stand for individual bands. The red solid line in a) represents a Lorentzian fit to the HM fibre at 1% strain using only a single G peak.

4. Position of G peak for a) HM fibre and b) P55 fibre. The rectangles represent the G peak positions when fitted as a single Lorentzian for $\theta_{in} = 0°$ and $\theta_{out} = 0°$. In a) the diamonds correspond to the $G^+$ (full) and $G^-$ (empty) positions for $\theta_{in} = 0°$. The straight lines represent the least-squares fits to the experimental data (solid lines for $\theta_{in} = 0°$ and dashed line for $\theta_{in} = 90°$- in the latter case the data points are not shown for the sake of clarity). The inserts show the FWHM of the respective G peaks when fitted as a single Lorentzian.

5. Master plot of $\partial \omega_G / \partial \varepsilon$ as a function of tensile modulus for the $G^+$ and $G^-$ of graphene and all sets of carbon fibres for laser $\theta_{in} = 0°$. The solid lines correspond to fits to experimental data (this work) for graphene whereas the dashed and dotted lines represent the analytical predictions. The solid and open triangles correspond to graphene G sub-bands values measured in this work and in Ref.[25], respectively. In black color, the solid square points correspond to Group A fibres, the solid circles correspond to Group B fibres, the open circles refer to Group C fibres[27], whereas the diamonds correspond to MPP carbon fibres reported in Ref.[18]. In red, data points for HM fibre (empty square), IM fibre (full circle), P25 (full diamond) and P55 (empty diamond) were acquired in the frame of this work. The red empty triangles show the G band splitting of the HM fibre and its projection onto the graphene average line. The inset shows the $\partial \omega_G / \partial \sigma$ least-squares line fits for Group A ($\partial \omega_G / \partial \sigma = 3$ cm$^{-1}$GPa$^{-1}$) and B (2.3 cm$^{-1}$GPa$^{-1}$), and MPP fibres (1.5 cm$^{-1}$GPa$^{-1}$).

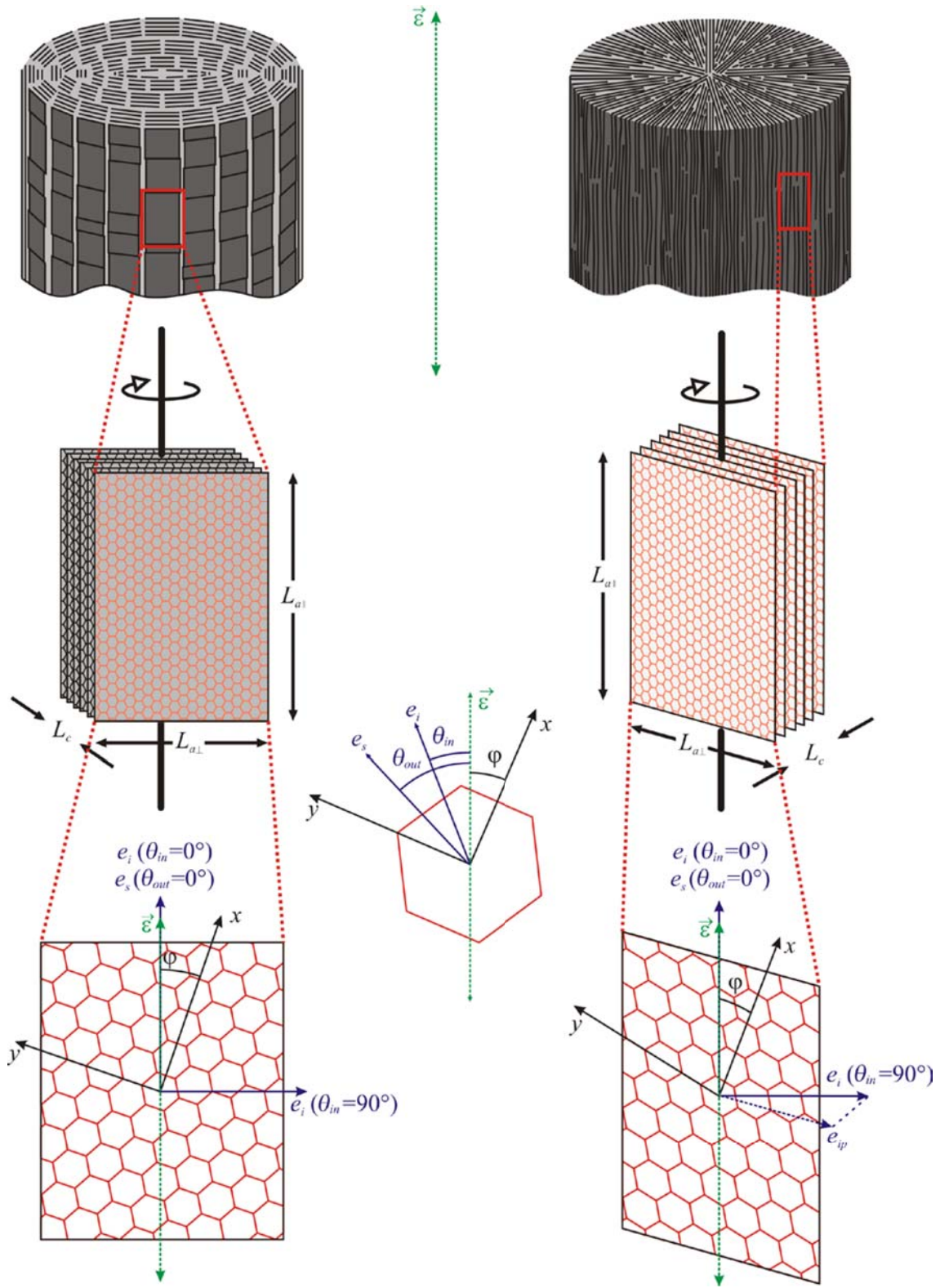

**Figure 1.** Schematic representation of scattering effects in the plane of graphene for onion-skin (left) and radial (right) carbon fibres. $\vec{\varepsilon}$ depicts the strain axis whereas $e_i$, and $e_s$ are *the* polarization directions of the incident and scattered light, respectively. $\theta_{in}$ and $\theta_{out}$ are the angles between the strain axis and the plane of the electric-field vector of incident and scattered light, respectively. The *x* axis has been taken perpendicular to the C-C bond. $\varphi$ is the angle between the strain axis and the *x* axis (i.e. orientation of the graphene lattice with respect to strain). $L_c$ is the carbon fibre crystallite thickness (i.e. number of graphene layers). $L_{a//}$ and $L_{a\perp}$ represent crystallite width in directions parallel and perpendicular to the fiber axis, respectively. $e_i$ is plotted in both directions, which were used in the experiments ($\theta_{in}$ = 0° and 90°). In the bottom right panel $e_{ip}$ designates the $e_i$ vector projected onto the graphene plane, which is rotated around a vertical axis at an arbitrary angle. $e_s$ is plotted only at $\theta_{out}$ = 0°.

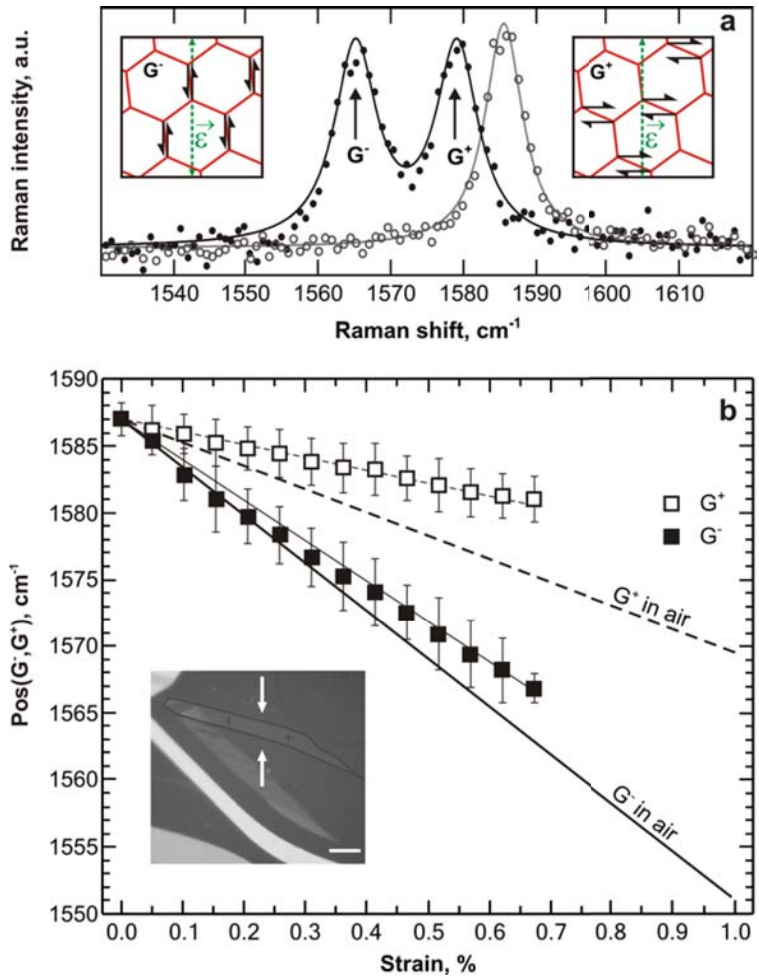

**Figure 2.** a) G peak Raman spectra of graphene at 0% (grey) and 1% (black) strain at 785 nm excitation. The original measurements are plotted as points. The solid curves are the best Lorentzian fits to the experimental spectra. b) $G^+$ (empty rectangles) and $G^-$ (full rectangles) peak positions for graphene embedded in polymer beam in tension. The data points are averages from five different locations on the flake. Solid lines represent values calculated for suspended graphene in air (see text). The inset shows an optical microphotograph of the measured flake. The scale bar on the inset is 10 μm and the arrows indicate the strain orientation.

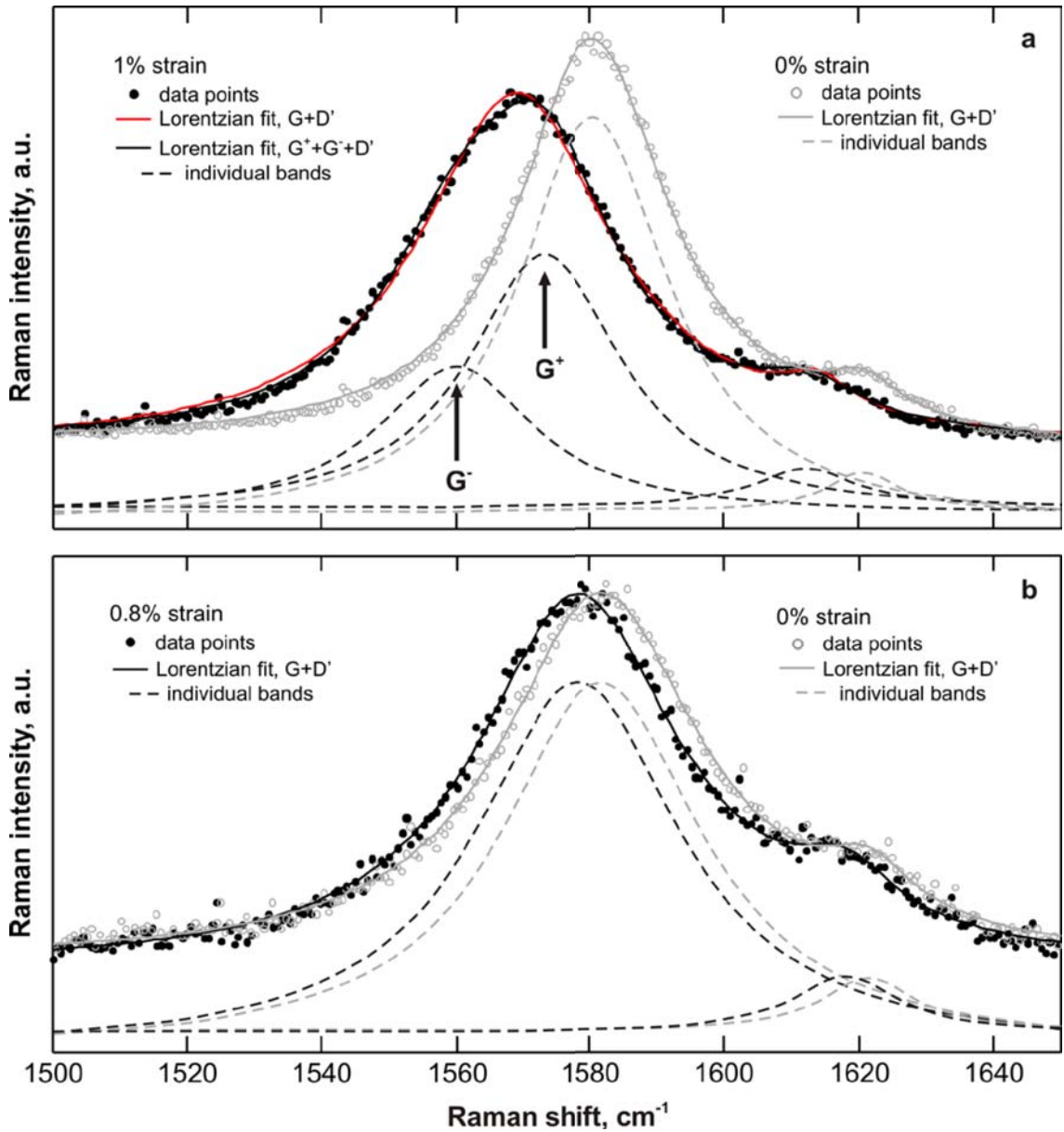

**Figure 3.** Raman G band region of a) HM fibre at 0 (G+D', grey) and 1% ($G^++G^-+D'$, black) strain and b) of P55 fibre at 0 (G+D', grey) and 0.8% (G+D', black) strain. The original measurements are plotted as points. The lines of respective colors represent best Lorentzian fits to the experimental spectra, where the dashed lines stand for individual bands. The red solid line in a) represents a Lorentzian fit to the HM fibre at 1% strain using only a single G peak.

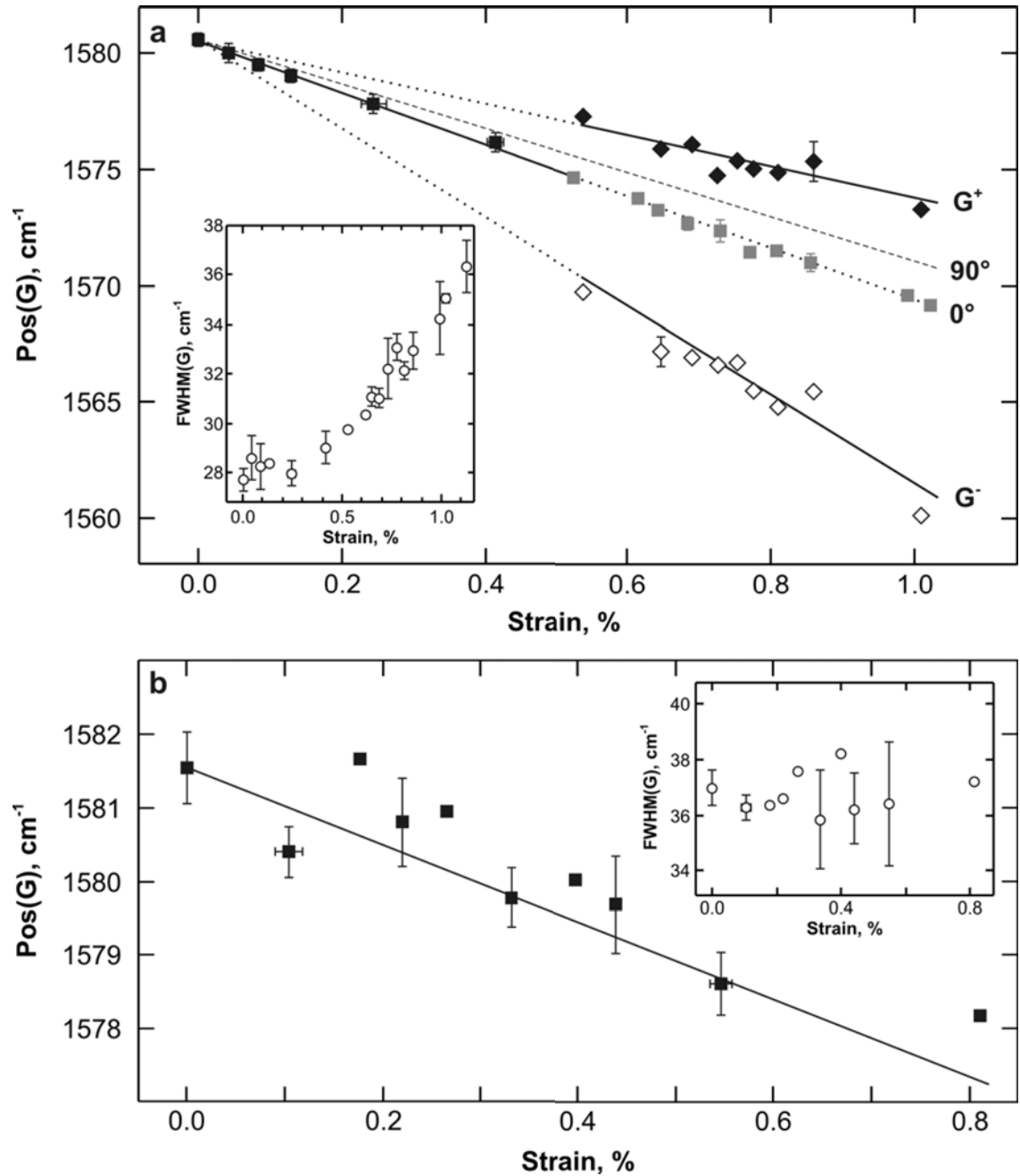

**Figure 4.** Position of G peak for a) HM fibre and b) P55 fibre. The rectangles represent the G peak positions when fitted as a single Lorentzian for $\theta_{in} = 0°$ and $\theta_{out} = 0°$. In a) the diamonds correspond to the $G^+$ (full) and $G^-$ (empty) positions for $\theta_{in} = 0°$. The straight lines represent the least-squares fits to the experimental data (solid lines for $\theta_{in} = 0°$ and dashed line for $\theta_{in} = 90°$ - in the latter case the data points are not shown for the sake of clarity). The inserts show the FWHM of the respective G peaks when fitted as a single Lorentzian.

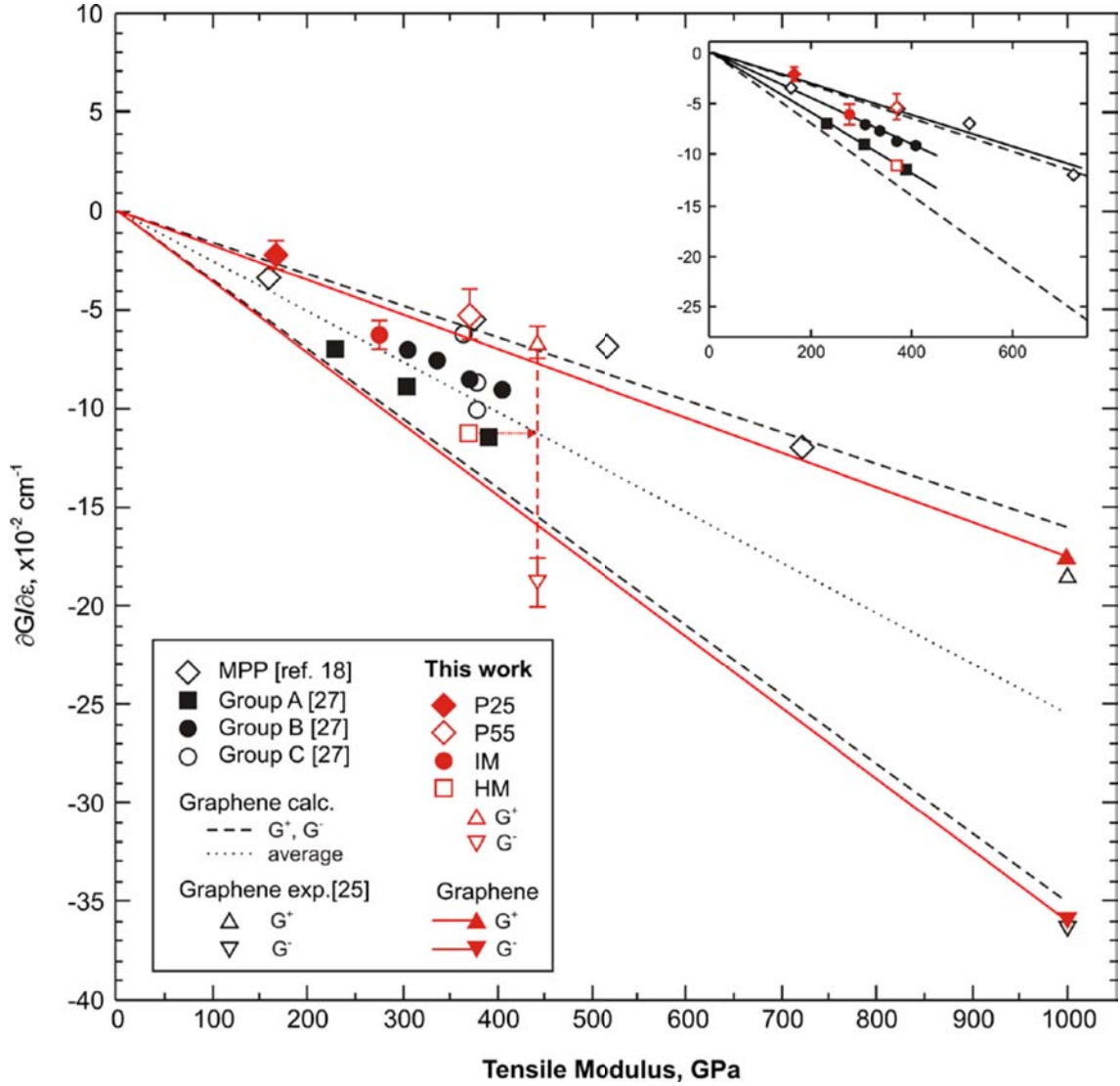

**Figure 5.** Master plot of $\partial \omega_G / \partial \varepsilon$ as a function of tensile modulus for the $G^+$ and $G^-$ of graphene and all sets of carbon fibres for laser $\theta_{in} = 0°$. The solid lines correspond to fits to experimental data (this work) for graphene whereas the dashed and dotted lines represent the analytical predictions. The solid and open triangles correspond to graphene G sub-bands values measured in this work and in Ref.[25], respectively. In black color, the solid square points correspond to Group A fibres, the solid circles correspond to Group B fibres, the open circles refer to Group C fibres[27], whereas the diamonds correspond to MPP carbon fibres reported in Ref.[18]. In red, data points for HM fibre (empty square), IM fibre (full circle), P25 (full diamond) and P55 (empty diamond) were acquired in the frame of this work. The red

empty triangles show the G band splitting of the HM fibre and its projection onto the graphene average line. The inset shows the $\partial \omega_G / \partial \sigma$ least-squares line fits for Group A ($\partial \omega_G / \partial \sigma = 3$ cm$^{-1}$GPa$^{-1}$) and B (2.3 cm$^{-1}$GPa$^{-1}$), and MPP fibres (1.5 cm$^{-1}$GPa$^{-1}$).